\documentclass[preprint, 3p, times,  onecolumn]{elsarticle} 
\usepackage{graphicx}
\usepackage{epsfig}
\usepackage{amsmath}
\usepackage{amsfonts}
\usepackage{amssymb}
\usepackage{dcolumn}
\usepackage{bm}
\usepackage{dcolumn}
\usepackage{color}
\usepackage{verbatim}
\usepackage{paralist}
\usepackage{amssymb}
\usepackage{graphicx}
\usepackage{extarrows}
\usepackage{url}
\usepackage{tabularx}
\usepackage{lineno}
\usepackage{etoolbox}
\usepackage[section]{placeins}
\usepackage[usestackEOL]{stackengine}
\usepackage{xspace}
\usepackage{xfrac}

\usepackage{lineno}

\journal{\texttt{arXiv.org}}

\def\lapprox{\mathrel{\mathop  {\hbox{\lower0.5ex\hbox{$\sim$}
\kern-1.1em\lower-0.7ex\hbox{$<$}}}}}
\def\gapprox{\mathrel{\mathop  {\hbox{\lower0.5ex\hbox{$\sim$}
\kern-1.1em\lower-0.7ex\hbox{$>$}}}}}

\newcommand{\bi}{$^{210}$Bi\xspace}
\newcommand{\po}{$^{210}$Po\xspace}
\newcommand{\pb}{$^{210}$Pb\xspace}
\newcommand{\be}{$^{7}$Be\xspace}
\newcommand{\pp}{$pp$\xspace}
\newcommand{\pep}{$pep$\xspace}
\newcommand{\cpd}{cpd/100t\xspace}

\usepackage{etoolbox}

\begin{document}

\begin{frontmatter}

\title {Independent determination of the Earth's orbital parameters with solar neutrinos in Borexino }


\author[Munchen]{S.~Appel}
\author[Juelich]{Z.~Bagdasarian\fnref{Berkeley}}
\author[Milano]{D.~Basilico}
\author[Milano]{G.~Bellini}
\author[PrincetonChemEng]{J.~Benziger}
\author[LNGS]{R.~Biondi}
\author[Milano]{B.~Caccianiga}
\author[Princeton]{F.~Calaprice}
\author[Genova]{A.~Caminata}
\author[Lomonosov]{A.~Chepurnov}
\author[Milano]{D.~D'Angelo}
\author[Peters]{A.~Derbin}
\author[LNGS]{A.~Di Giacinto}
\author[LNGS]{V.~Di Marcello}
\author[Princeton]{X.F.~Ding}
\author[Princeton]{A.~Di Ludovico} 
\author[Genova]{L.~Di Noto}
\author[Peters]{I.~Drachnev}
\author[APC]{D.~Franco}
\author[Princeton,GSSI]{C.~Galbiati}
\author[LNGS]{C.~Ghiano}
\author[Milano]{M.~Giammarchi}
\author[Princeton]{A.~Goretti\fnref{LNGSG}}
\author[Juelich,RWTH]{A.S.~G\"ottel}
\author[Lomonosov,Dubna]{M.~Gromov}
\author[Mainz]{D.~Guffanti\fnref{Bicocca}}
\author[LNGS]{Aldo~Ianni}
\author[Princeton]{Andrea~Ianni}
\author[Krakow]{A.~Jany}
\author[Kiev]{V.~Kobychev}
\author[London,Atomki]{G.~Korga}
\author[Juelich,RWTH]{S.~Kumaran}
\author[LNGS]{M.~Laubenstein}
\author[Kurchatov,Kurchatovb]{E.~Litvinovich}
\author[Milano]{P.~Lombardi}
\author[Peters]{I.~Lomskaya}
\author[Juelich,RWTH]{L.~Ludhova}
\author[Kurchatov]{G.~Lukyanchenko}
\author[Kurchatov,Kurchatovb]{I.~Machulin}
\author[Mainz]{J.~Martyn}
\author[Milano]{E.~Meroni}
\author[Milano]{L.~Miramonti}
\author[Krakow]{M.~Misiaszek}
\author[Peters]{V.~Muratova}
\author[Kurchatov,Kurchatovb]{R.~Nugmanov}
\author[Munchen]{L.~Oberauer}
\author[Mainz]{V.~Orekhov}
\author[Perugia]{F.~Ortica}
\author[Genova]{M.~Pallavicini}
\author[Juelich,RWTH]{L.~Pelicci}
\author[Juelich]{\"O.~Penek}
\author[Princeton]{L.~Pietrofaccia}
\author[Peters]{N.~Pilipenko}
\author[UMass]{A.~Pocar}
\author[Kurchatov]{G.~Raikov}
\author[LNGS]{M.T.~Ranalli}
\author[Milano]{G.~Ranucci}
\author[LNGS]{A.~Razeto}
\author[Milano]{A.~Re}
\author[Juelich,RWTH]{M.~Redchuk\fnref{Padova}}
\author[LNGS]{N.~Rossi}
\author[Munchen]{S.~Sch\"onert}
\author[Peters]{D.~Semenov}
\author[Juelich]{G.~Settanta\fnref{ISPRA}}
\author[Kurchatov,Kurchatovb]{M.~Skorokhvatov}
\author[Juelich,RWTH]{A.~Singhal}
\author[Dubna]{O.~Smirnov}
\author[Dubna]{A.~Sotnikov}
\author[LNGS]{R.~Tartaglia}
\author[Genova]{G.~Testera}
\author[Peters]{E.~Unzhakov}
\author[Dubna]{A.~Vishneva}
\author[Virginia]{R.B.~Vogelaar}
\author[Munchen]{F.~von~Feilitzsch}
\author[Krakow]{M.~Wojcik}
\author[Mainz]{M.~Wurm}
\author[Genova]{S.~Zavatarelli}
\author[Dresda]{K.~Zuber}
\author[Krakow]{G.~Zuzel}

\fntext[Berkeley]{Present address: University of California, Berkeley, Department of Physics, CA 94720, Berkeley, USA}
\fntext[LNGSG]{Present address: INFN Laboratori Nazionali del Gran Sasso, 67010 Assergi (AQ), Italy}
\fntext[Padova]{Present address: Dipartimento di Fisica e Astronomia dell'Universit\`a di Padova and INFN Sezione di
Padova, Padova, Italy}
\fntext[ISPRA]{Present address: Istituto Superiore per la Protezione e la Ricerca Ambientale, 00144 Roma, Italy}
\fntext[Bicocca]{Present address: Dipartimento di Fisica, Università degli Studi e INFN Milano-Bicocca, 20126 Milano, Italy}

\address{\bf{The Borexino Collaboration \\  (\texttt{spokesperson-borex@lngs.infn.it})}}

\address[APC]{AstroParticule et Cosmologie, Universit\'e Paris Diderot, CNRS/IN2P3, CEA/IRFU, Observatoire de Paris, Sorbonne Paris Cit\'e, 75205 Paris Cedex 13, France}
\address[Dubna]{Joint Institute for Nuclear Research, 141980 Dubna, Russia}
\address[Genova]{Dipartimento di Fisica, Universit\`a degli Studi e INFN, 16146 Genova, Italy}
\address[Krakow]{M.~Smoluchowski Institute of Physics, Jagiellonian University, 30348 Krakow, Poland}
\address[Kiev]{Institute for Nuclear Research of NAS Ukraine,
03028 Kyiv, Ukraine}
\address[Kurchatov]{National Research Centre Kurchatov Institute, 123182 Moscow, Russia}
\address[Kurchatovb]{ National Research Nuclear University MEPhI (Moscow Engineering Physics Institute), 115409 Moscow, Russia}
\address[LNGS]{INFN Laboratori Nazionali del Gran Sasso, 67010 Assergi (AQ), Italy}
\address[Milano]{Dipartimento di Fisica, Universit\`a degli Studi e INFN, 20133 Milano, Italy}
\address[Perugia]{Dipartimento di Chimica, Biologia e Biotecnologie, Universit\`a degli Studi e INFN, 06123 Perugia, Italy}
\address[Peters]{St. Petersburg Nuclear Physics Institute NRC Kurchatov Institute, 188350 Gatchina, Russia}
\address[Princeton]{Physics Department, Princeton University, Princeton, NJ 08544, USA}
\address[PrincetonChemEng]{Chemical Engineering Department, Princeton University, Princeton, NJ 08544, USA}
\address[UMass]{Amherst Center for Fundamental Interactions and Physics Department, University of Massachusetts, Amherst, MA 01003, USA}
\address[Virginia]{Physics Department, Virginia Polytechnic Institute and State University, Blacksburg, VA 24061, USA}
\address[Munchen]{Physik-Department, Technische Universit\"at  M\"unchen, 85748 Garching, Germany}
\address[Lomonosov]{Lomonosov Moscow State University Skobeltsyn Institute of Nuclear Physics, 119234 Moscow, Russia}
\address[GSSI]{Gran Sasso Science Institute, 67100 L'Aquila, Italy}
\address[Dresda]{Department of Physics, Technische Universit\"at Dresden, 01062 Dresden, Germany}
\address[Mainz]{Institute of Physics and Excellence Cluster PRISMA+, Johannes Gutenberg-Universit\"at Mainz, 55099 Mainz, Germany}
\address[Juelich]{Institut f\"ur Kernphysik, Forschungszentrum J\"ulich, 52425 J\"ulich, Germany}
\address[RWTH]{III. Physikalisches Institut B, RWTH Aachen University, 52062 Aachen, Germany}
\address[London]{Department of Physics, Royal Holloway University of London, Egham, Surrey,TW20 0EX, UK}
\address[Atomki]{Institute of Nuclear Research (Atomki), Debrecen, Hungary}

\begin{abstract}

Since the beginning of 2012, the Borexino collaboration has been reporting precision measurements of the solar neutrino fluxes, emitted in  the proton--proton chain and in the Carbon-Nitrogen-Oxygen cycle. The experimental sensitivity achieved in Phase-II and Phase-III of the Borexino data taking made it possible to detect the annual modulation of the solar neutrino interaction rate  due to the eccentricity of Earth's orbit, with a statistical significance greater than 5$\sigma$. This is
the first precise measurement of the Earth's orbital parameters based solely on solar neutrinos and an additional signature of the solar origin of the Borexino signal. The complete periodogram of the time series of the Borexino solar neutrino detection rate is also reported, exploring frequencies between one cycle/year and one cycle/day. No other significant modulation frequencies are found. 
The present results were uniquely made possible by Borexino's decade-long high-precision solar neutrino detection.

\end{abstract}


\end{frontmatter}
\twocolumn
\sloppy

\section*{Introduction}

The motivation for this measurement has a rich historical context, often unnoticed by the modern reader.
The first heliocentric hypothesis, i.e. the astronomical model in which the Earth and planets revolve around the Sun at the center of the Universe,  was proposed by Aristarchus of Samos in the third century BC, in order to simplify the complex system of planet retrograde motions, due to the apparent picture of considering the Earth at the center of the cosmos. This early brilliant intuition was definitely overwritten a few centuries later by the geocentric model by Claudios Ptolemy who reported in his famous treatise \emph{The Almagest} a full description of the planet motions as seen from the Earth, laying the foundation of the long-lasting Medieval conception of the Universe. In spite of the very advanced level of the ancient Greek science reached during the Hellenistic age, it is not clear whether the elliptical nature of the Earth orbit was known. Some hypotheses in favor has been put forward, because the curve ellipse was largely described in \emph{The Conics} of Apollonius of Perga and Sun-Earth changing distance was known \cite{bib:vlad}, but there is lack of certain historical sources. For further details, see \cite{bib:russo}.

\begin{figure}[h!]
    \centering
    \includegraphics[width=0.94\columnwidth]{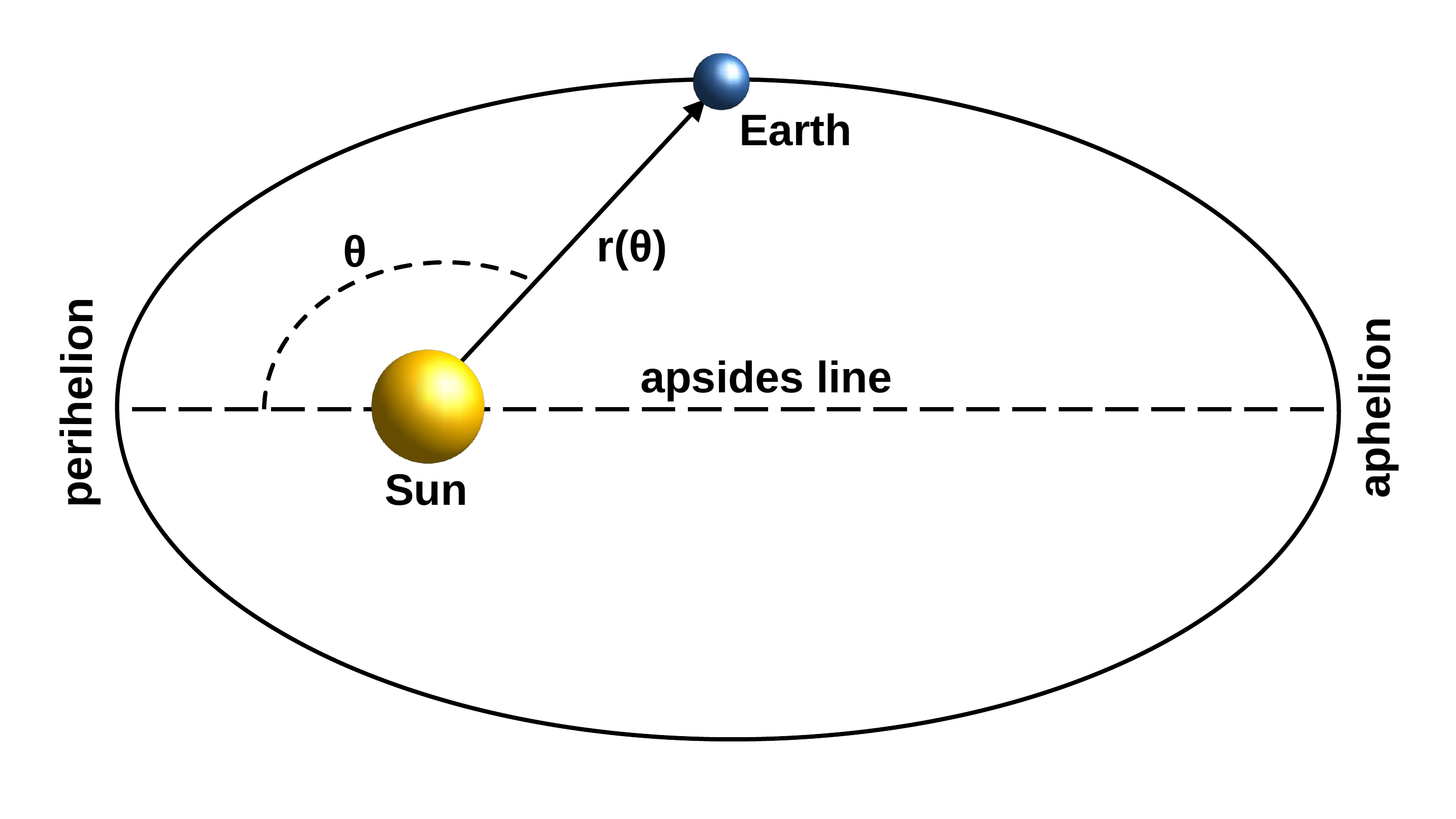}
    \caption{Earth's orbit around the Sun with parameters of interest. The Earth revolves around the Sun keeping the distance $r(\theta)$, where $\theta$ is the angle with respect to the perihelion in polar coordinates. The eccentricity $\epsilon$ is defined as the ratio between the difference and the sum of the aphelion and perihelion.}
    \label{fig:orbit}
\end{figure}

Many centuries later, as a consequence of the art and science flowering of the Renaissance period, the heliocentric model came again into existence in 1543 thanks to Nicholas Copernicus, who first redrew the heliocentric model approximating with circles the planets motions in his famous \emph{De Rivolutionibus Orbium Coelestium}. The scientific and philosophical dispute was really intense at that time, and very well summarized in the \emph{Dialogue Concerning the Two Chief World Systems} published by Galileo Galilei in 1632. Later on, Johannes Kepler, taking advantage of the high precision astronomical measurement of his mentor Tycho Brahe,  improved the heliocentric model through his Three Laws of Planetary Motions in which for the first time the elliptical nature of the planet orbit, including the Earth, were accurately stated
(the first two laws are in \emph{Astronomia Nova}, published in 1609, and the last in \emph{Harmonices Mundi}, published in 1619). In particular the First Law states that all planetary orbits are elliptical and Sun occupies one of its two foci.
Those important pieces of information allowed Isaac Newton to formulate the Law of Universal Gravitation in his work \emph{Philosophiae Naturalis Principia Mathematica} published in 1687, so far considered as one of the greatest achievements of human thought. For further details, see  \cite{bib:segre}. The eccentricity $\epsilon$, defined as the ratio between the difference and the sum of the Earth's aphelion and perihelion (see Fig. \ref{fig:orbit}), quoted in the \emph{Principia} is $16\sfrac{7}{8}$ over 1000 parts, i.e. $\epsilon = 0.0169$ in modern numbers, very close to the current astronomical measurement \cite{bib:modern_e, bib:nasa_e}, rounded to 0.0167 for the purpose of this work.

In this work, the first precise measurement of the Earth's orbit eccentricity exploiting the variation of the solar neutrino flux produced in the Sun's core and detected by Borexino on the Earth, caused by the Sun-Earth distance change as a function of time due to the non-circular shape of the orbit, is reported. Since neutrinos can travel through the Earth and then detected 24 hours a day, the flux change depends only on the inverse-square of the Earth-Sun distance. Using the polar coordinate, that distance, can be written as
\begin{equation}
    r(\theta) = \frac{\bar{r} (1-\epsilon^2)}{1 + \epsilon \cos(\theta)}\,
\end{equation}
where $\bar{r}$ is the average of the apsides and $\theta$ is the polar angle with respect to the perihelion. 
Since $\epsilon \ll 1$, the solar neutrino flux, produced by the Sun as $\Phi_0$, and hitting the Earth at the time $t$, can be approximated coherently with Kepler's Second Law by
\begin{equation}
    \Phi(t) \approx \frac{\Phi_0}{\bar{r}^2}\left[1 + 2 \epsilon \cos(\omega_y(t - t_0))\right] + \mathcal{O}(\epsilon^2). 
    \label{eq:flux}
\end{equation}
Where $\omega_y = 2\pi/T_y$ is the Earth's average angular velocity over a year $T_y$ and $t_0$ is the phase that can be chosen at the perihelion (usually falling on the first days of January). 
The expected percent amplitude variation is of order $A = 2\epsilon \approx 3.37\%$.
The result presented in this work is important for two main reasons: first, because it provides an independent proof of the Kepler's first law, that has its own fascinating philosophical aspect; second, because it proves the unprecedented level of precision and stability of solar neutrino detection achieved by Borexino. Furthermore, profiting off the very stable time series, supported by the annual modulation detection, the full periodogram of the solar neutrino time series, exploring frequencies up to one cycle/day, is hereby reported. That analysis has its own importance for other signals of interest for the solar physics and non-standard neutrino interaction (NSI), as Sun's rotation or Earth's day-night asymmetry. 

In Sec. \ref{sec:bx} the main results of Borexino concerning solar neutrino physics are summarized.  In Sec. \ref{sec:data} data selection criteria are described. In Sec. \ref{sec:gls} the \emph{Generalized Lomb-Scargle} for the frequency analysis is reviewed and applied to the Borexino time series for the periodic signal quest. In Sec. \ref{sec:epsilon} the eccentricity and other Earth's orbit parameters are reported and compared to previous solar neutrino experiments. Finally, in Sec. \ref{sec:freq} the search for other possible modulated signals is largely detailed.


\section{The Borexino detector}
\label{sec:bx}

Borexino is the only solar neutrino experiment  
able of reconstructing the position and the energy of each event in real-time with an analysis energy threshold of $E_{th}\approx150$ keV, thanks to the ultra-low level of its radioactive background.

Borexino is located in the Hall C of Laboratori Nazionali Gran Sasso (LNGS-INFN) \cite{bib:lngs}. The detector is made of concentric shells with increasing radiopurity (see e.g. Ref. \cite{bib:tech}): the innermost core, enclosed in a 125 $\mu$m thick ultra-pure nylon vessel of radius 4.25 m, is made of about 280 tons of liquid scintillator (1,2,4-Trimethylbenzene  with 1.5 g/l of PPO wavelength shifter). The active core is contained in a stainless steel sphere (SSS) filled up with $\sim 1000$ tons of buffer liquid (1,2,4-Trimethylbenzene  with DMP quencher), whose internal surface is instrumented with more than 2000 PMTs for detecting the scintillation light. Finally, the SSS is located inside a 2000 tonne water Cherenkov detector, equipped with about 200 PMTs. 
Thanks to an intense calibration campaign carried out in 2010, the Borexino detector is able to reconstruct the event position with an accuracy of $\sim 10$\,cm (at 1 MeV)  and with energy resolution of about $\sigma(E) / E = 5\%/\sqrt(E/[MeV])$ \cite{bib:calib}. 

The Borexino data-set is traditionally divided in three Phases, spaced out by hardware milestones: Phase-I, from mid-2007 to beginning of 2010, ends with the calibration campaign, in which the first measurement of the \be solar neutrino interaction rate \cite{bib:be7-1, bib:be7-2, bib:be7-3} has been performed; Phase-II, from the-beginning of 2012 to mid-2016, starts after an intense purification campaign, based on water extraction, with unprecedented suppression of the radioactive contaminants, in which the first evidence of the \pep neutrinos \cite{bib:pep} and a 10\% measurement of the \pp neutrinos \cite{bib:pp} has been published, later updated in the solar neutrino comprehensive analysis \cite{bib:global, bib:nusol, bib:b8}; Phase-III, from mid-2016 (end of the thermal insulation installation) to October 3rd 2021 (beginning of the detector decommissioning).
In the first part of Phase-III the first detection of the CNO neutrinos \cite{bib:cno} has been performed. Table \ref{tab:solars} summarizes the most important results concerning solar neutrinos interaction rates measured by Borexino.
\begin{table}[h!]
\small
\begin{center}
\begin{tabular}{|lll|} 
\hline  
Species & Rate [cpd/100t] & Flux [cm$^{-2}$ s$^{-1}$ ] \\ \hline
\emph{pp}
& $(134 \pm 10)^{+6}_{-10}  $ &  $(6.1 \pm 0.5)_{-0.5}^{+0.3} \cdot 10^{10}$   \\
$^7$Be & $(48.3 \pm 1.1)^{+0.4}_{-0.7}$ &  $(4.99 \pm 0.11)_{-0.08}^{+0.06} \cdot 10^9$ \\
\emph{pep}  (HZ)   & $(2.7 \pm 0.4)^{+0.1}_{-0.2} $  & $(1.3 \pm 0.3)_{0.1}^{+0.1} \cdot 10^8$ \\ 
$^8$B& $ 0.223_{-0.022}^{+0.021}$ & $ 5.68_{-0.44}^{+0.42} \cdot 10^6$ \\
CNO    & $7.2_{-1.7}^{+3.0}$    &  $7.0_{-2.0}^{+3.0} \cdot 10^8$ \\
\emph{hep} & $< 0.002$ (90\% CL) & $<1.8 \cdot 10^5$ (90\% CL)   
\\ \hline 
\end{tabular}
\end{center}
\caption{Solar neutrino interaction rates and fluxes measured by Borexino.  Rates are reported in counts per day per 100 tonne (\cpd), while fluxes are reported in cm$^{-2}$s$^{-1}$. The $pep$ rate is reported under the \emph{high-metallicity} hypothesis (HZ), see \cite{bib:global} for further details.}
\end{table} \label{tab:solars}
Thanks to its unprecedented radio-purity level, Borexino has also set important limits on rare processes (see e.g., \cite{bib:elec, bib:nsi, bib:astro-nu, bib:magnet, bib:sterile} and performed other neutrino physics studies, as e.g. geo-neutrino detection (for review, see e.g. \cite{bib:geonu}). 
\begin{figure}[h!]
    \centering
    \includegraphics[width=0.94\columnwidth]{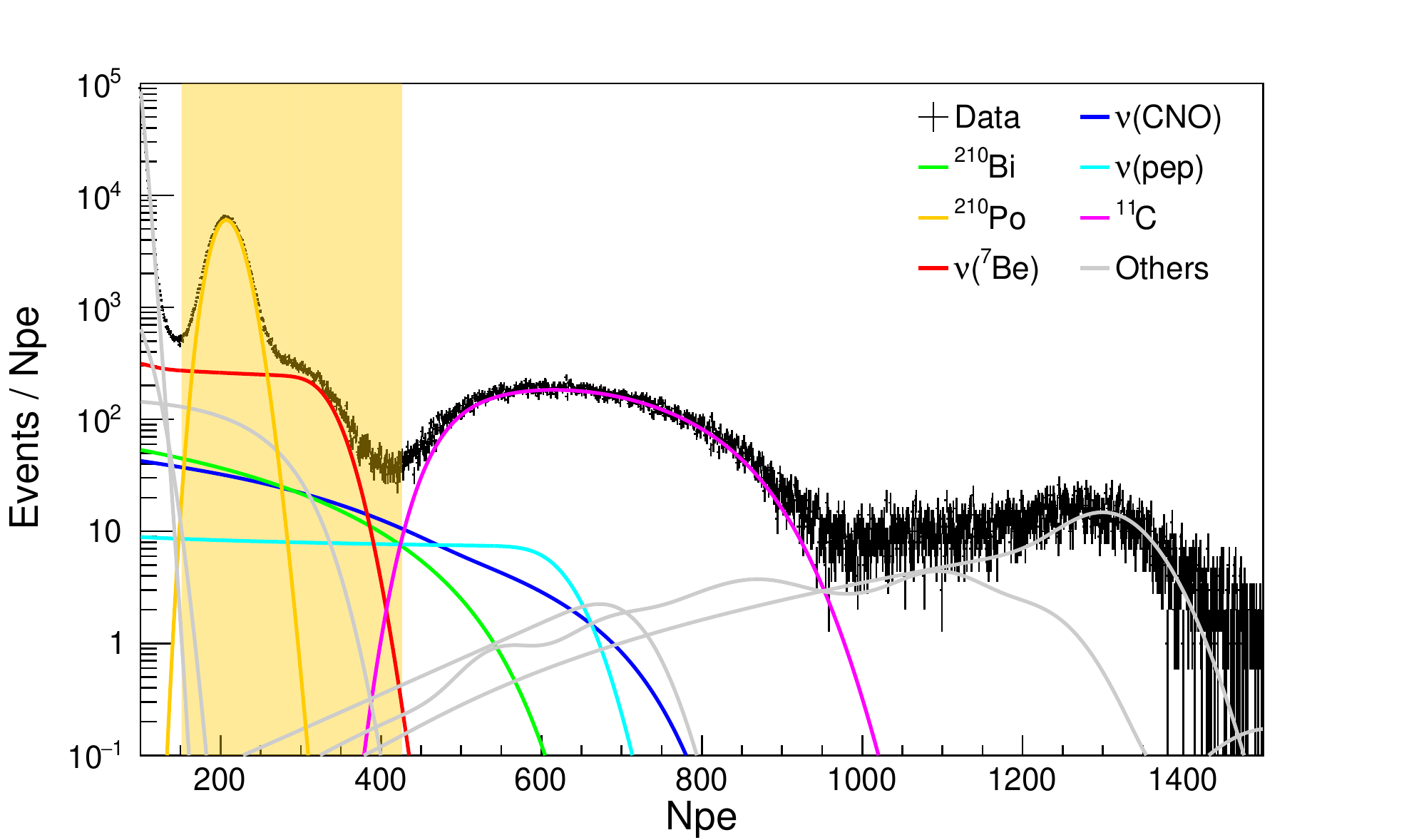}
    \caption{Borexino energy spectrum as a function of the energy estimator ``geometry-normalized \texttt{Npe}''. Typical contribution to the spectrum in terms of solar neutrino electron scattering and $\beta$-like background are reported in the legend. The vertical yellow band shows the constant energy window region used for building up the time series of the total rate.}
    \label{fig:bxspec}
\end{figure}

The $\beta$-like event selection in Borexino for neutrino candidate identification is described in details in \cite{bib:long}: the basic idea is to select point-like events in the innermost part of the Inner Vessel, avoiding cosmogenic background induced by muons crossing the scintillator and limiting the internal contamination from radioactive decays.
It is worth mentioning that the evidence of the annual modulation of the solar neutrino flux due to the Earth's orbit eccentricity in Borexino was already reported in Phase-I (3 years, $\approx 3\sigma$ level \cite{bib:long}) and the first part of Phase-II (5 years, $3.5 \sigma$ level \cite{bib:seas}). The previous analysis was performed using the Lomb-Scargle method \cite{Lomb:1976wy,Scargle:1982bw}, the Empirical Mode Decomposition approach \cite{bib:emd}, and the standard least square sinusoidal fit.  

\section{Data selection}
\label{sec:data}

The time analysis of the solar neutrino interaction rate in Borexino is performed through signal processing techniques, applied to the time series of the total event rate in a fixed energy window. The energy estimator is the number of photo-electrons after an event position-dependent correction for the spherical geometry 
(``geometry-normalized \texttt{Npe}''), see Ref.~\cite{bib:long}. To maximize the signal-to-background for solar neutrinos,
a region of interest (RoI) in the energy spectrum 
is chosen according to the following Figure of Merit (FoM):
\begin{equation}
FoM\,(\Delta E) = \frac{R_S}{\sigma(R_{tot})},    
\end{equation}
where $R_S$ is the solar neutrino rate in the energy interval $\Delta E$ and $\sigma(R_{tot})$ is the width of the distribution of the total rate $R_{tot}$ in said energy range. The chosen energy RoI is 150-428 \texttt{Npe}, corresponding to 300-827 keV, is highlighted in yellow in Fig.~\ref{fig:bxspec}.  
The mono-energetic peak of quenched \po alpha events 
is drastically reduced via high efficiency pulse shape discrimination based on \emph{multi-layer perceptron} (MLP)~\cite{bib:nusol, bib:mlp}. What is left in the energy RoI is mostly from the mono-energetic \be solar neutrinos, whose electron scattering is characterized by a typical Compton-like shoulder, and a subdominant contribution from \pep and CNO neutrinos. Backgrounds contributing to this region include: \bi and $^{85}$Kr $\beta$ decays, a small $\gamma$-ray contribution from $^{40}$K, $^{214}$Bi, $^{208}$Tl external to the fiducial volume, and cosmogenic $^{11}$C $\beta^+$ decays. The Borexino $\beta$-like spectrum is reported in full detail in Ref.~\cite{bib:long}.  

The analysis is performed on data taken between December 11$^{th}$ 2011 and October 3$^{rd}$ 2021, when detector decommissioning operations began. This period of almost 10 years, includes the Phase-II and Phase-III data used in the aforementioned analyses, extended to include data collected after 
February 2020. Data are selected in a spherical fiducial volume (FV) of 3 meter diameter (about 100 tonnes). This volume is larger than the typical FVs used by Borexino in previous solar neutrino analyses 
and unavoidably includes more background, notably from $\gamma$-rays from external detector elements such as the Nylon Vessel and its supporting structures, and the PMTs. These contributions are, however, basically constant or very slowly varying and are referred to as \emph{secular} variations in the following.

We do not include the Phase-I data in the present analysis
because of the high content of \po and the drastically different contribution of other backgrounds, such as \bi and $^{85}$Kr, that were significantly reduced by the scintillator purification campaign at the end of Phase-I. 
Due to the difference in detector conditions before and after purification, the inclusion of Phase-I in the current analysis would not significantly improve the measurement.

The secular variation in the Borexino time series, clearly visible in Fig. \ref{fig:d30} (\emph{Top}), has a different origin. The initial fast decay is attributed to leakage of alpha events through the MLP pulse shape discrimination due to its $\sim$ 1\% tagging inefficiency.
Alpha particles originate almost exclusively from the \po decay. At the beginning of Phase-II, 
the out-of-equilibrium \po present in the scintillator bulk is measured at $\sim 1400$ \cpd 
After three \po life times ($\sim 600$ days), this decay component becomes sub-dominant, but a residual \po migration from the Nylon Vessel is observed for the entire period with a rate of about 30 \cpd in the FV. Details about the \po migration and its importance in the CNO analysis is described in Ref.~\cite{bib:cno}. Finally, an almost constant component of \po, supported by the decays of parent \bi in secular equilibrium with trace amounts of long-lived \pb ($\tau\sim 32$ years) in the scintillator is also present and hardly visible.

A further secular variation comes from the \bi initial non-uniformity (Phase-II). Following the last stage of scintillator purification via water extraction, a more radio pure scintillator was introduced from the top of the detector 
which generated a top--bottom asymmetry in the \pb activity. This asymmetry gradually smoothed out through convective motions in the scintillator and was gone when
the thermal insulation of the detector began to enable the measurement of CNO neutrinos. 
Another possible secular variation of the total rate in the RoI is ascribable to $^{85}$Kr ($\tau \approx 10$ y). This contaminant was drastically reduced 
by the scintillator purification from an initial activity of $\sim 30$ \cpd to $\lesssim 5$ \cpd. The $^{85}$Kr rate can be independently quantified through its 0.43\% BR $\beta$--$\gamma$ time correlated decay mode. 
The extremely low $^{85}$Kr concentration and the small branching ratio of this decay mode do not allow a clear determination of whether there is an incresing or decreasing $^{85}$Kr trend during Phase-II and beyond.
An increase could be due to migration from the Outer Buffer fluid through the Nylon Vessel membrane into the scintillator. In either case, this contribution is expected to be monotonic in time and easily removable by the data \emph{detrending} procedure used to eliminate overall long-term trends and emphasise higher frequency components.   

The last important contribution to the secular variation is the slow deformation of the Nylon Vessel over time. The shape of the vessel is precisely monitored through the  background contamination present on its surface~\cite{bib:long}. Considering a standard polar coordinate system, the distance of the vessel surface from the center $d(\theta,\phi)$ deviates slightly from its nominal value ($r_0=4.25$ m). It is observed that in the period of this analysis, the vessel displays a slow, monotonic deformation along $\theta$ while preserving azimuthal ($\phi$) symmetry. Such a slow deformation can subtly affect the contribution from external $\gamma$-ray background
in the RoI. 
The detrending procedure used to filter out these secular contributions is discussed in the next Section. 

\section{Frequency analysis}
\label{sec:gls}
The Lomb-Scargle (LS) periodogram is a standard generalization of the Fourier transform for the spectral analysis of time series consisting of unequally spaced data.  Its statistical properties are valid under the assumption of time series affected by Gaussian fluctuations. This is reasonably accepted for event rates containing more than 30 events per time interval, while lower statistics are properly described by Poissonian fluctuations.  

\begin{figure}[h!]
    \centering
    \includegraphics[width=0.94\columnwidth]{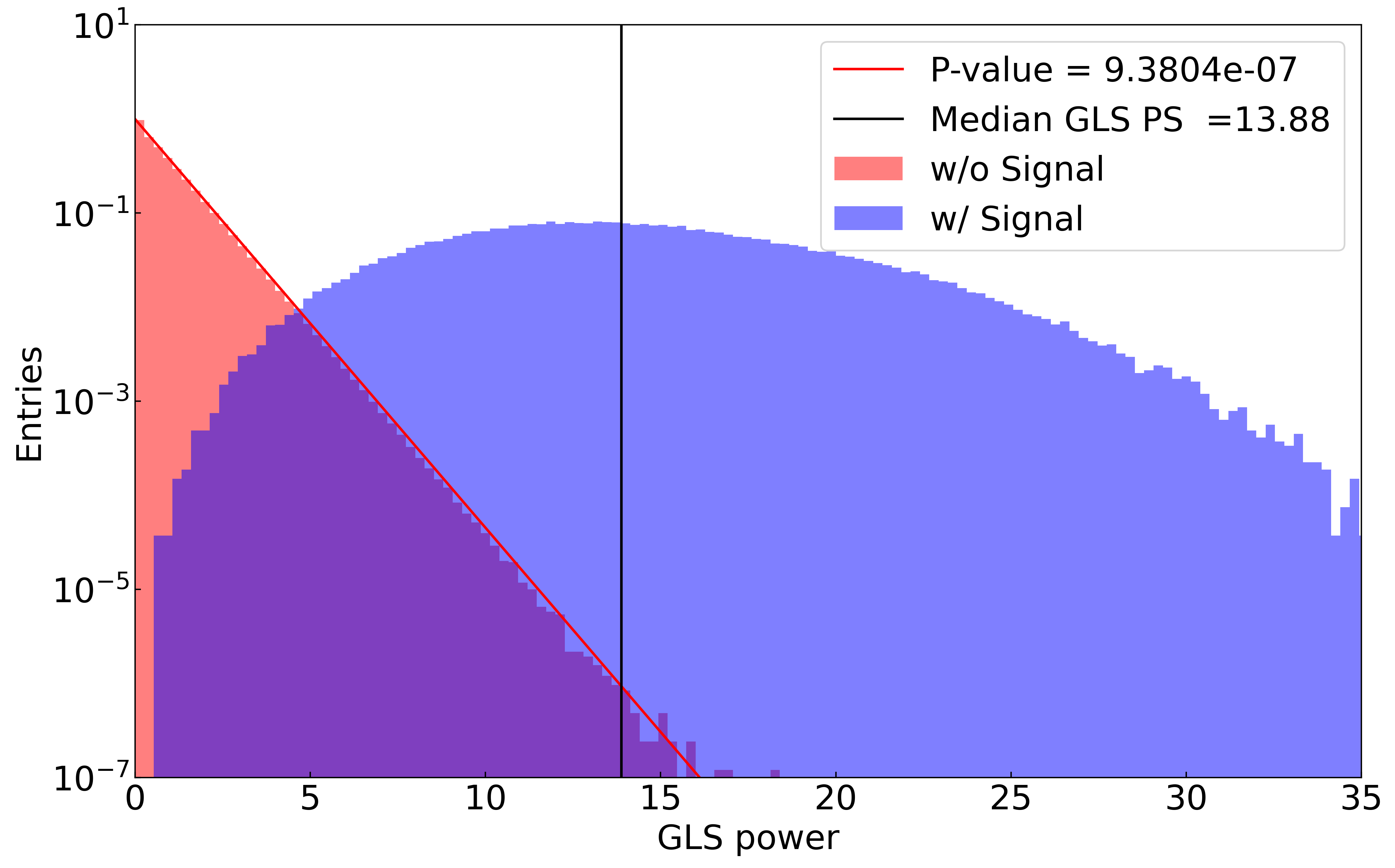}
    \caption{Median sensitivity for the detection of the annual modulation in the 30 day binning of the Borexino time series.}
    \label{fig:sens}
\end{figure}

\begin{figure*}[h!]
    \centering
    \includegraphics[width=0.94\textwidth]{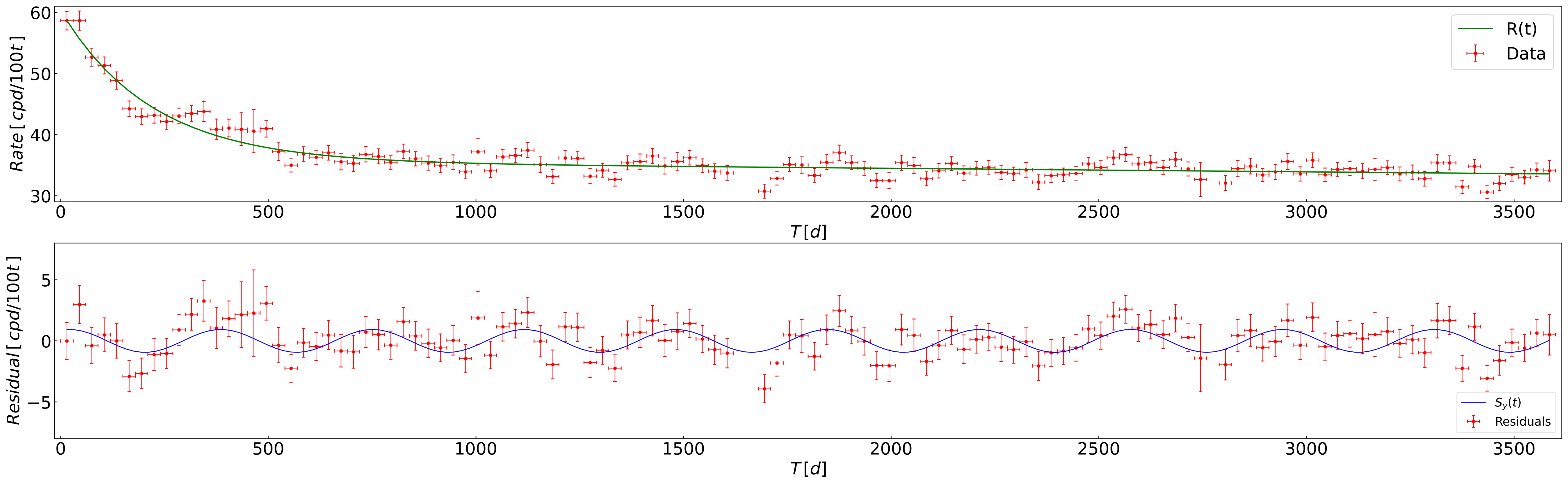}
    \caption{\emph{Top}: Full Borexino rate time series (Phase-II and Phase-III) in the RoI fitted to the trend model $R(t)$ to remove secular components. The rate in \cpd is binned in time intervals of 30 days. The time axis is reported in days since 12:00 AM of December 11\textsuperscript{th} 2011, in UTC time. \emph{Bottom}: Residuals of the time series with respect to the trend model $R(t)$. The blue sinusoidal best fit of the residual rate indicates the presence of a significant annually modulated signal.}
    \label{fig:d30}
\end{figure*}

As shown in \cite{Ranucci:2005ep, Sturrock:1997gp, Ranucci:2006rz}, the LS method can be generalized via a likelihood approach to what is referred to as the Generalized Lomb-Scargle (GLS) method. Searching for an annual modulation does not strictly require a GLS approach since, in our case, one-month data bins feature a sufficiently high statistics. However, when searching for higher frequency signals, {\it e.g.} a day-night asymmetry in the neutrino rate, the Nyquist theorem requires a time binning shorter than half of the inverse of the investigated frequency. This inevitably pushes the event rate per bin in the Poissonian regime. For this reason, all analyses presented below rely on the GLS approach.

The expected number of solar neutrinos detected in the $i$-th time interval $t_i$ is given by:
\begin{equation}
    \mu_i = \mu_{trend}(t_i) \left[1 + A \cos \left(2\pi \nu( t_i + \phi)\right) \right] ,
\end{equation}
where $\mu_{trend}(t)$ is the detrending function, $A$, $\nu$, and $\phi$ are the relative amplitude, frequency, and phase of the neutrino flux modulation, respectively. For Poissonian statistics, one can build the following likelihood function:
\begin{equation}
  L =   \prod^N_i  \frac{\mu_i^{n_i} ~ e^{-\mu_i} }{n_i!}  \label{eq:like}
\end{equation}
According to  Wilks's theorem \cite{Wilks}, the generalized likelihood ratio (GLR) can be written as:
\begin{equation}
     GLR(\nu) = \frac{ \displaystyle \prod^N_i  \frac{\mu_{trend}^{n_i} ~ e^{-\mu_{trend}} }{n_i!} }{ \displaystyle \max_{A,\phi} \prod^N_i  \frac{\mu_i^{n_i} ~ e^{-\mu_i} }{n_i!} }
\end{equation}
The same theorem states that $S = -\ln(GLR)$  is exponentially distributed as $e^{-S}$ under the null hypothesis. It is also by definition the likelihood spectrum of the signal, sharing the same properties of the LS periodogram $S = \Delta \chi^2$, which in turn corresponds to the Fourier power spectrum when the time series is normalized to its RMS.
In other words, the LS is a special case of the GLS method when the errors have a Gaussian distribution~\cite{Ranucci:2005ep}.
To use binned data normalized to different live times in the standard Borexino unit of \cpd, one can recast Eq. (\ref{eq:like}) into:
\begin{equation}
     L = \prod^N_{i=0} \frac{y_i^{x_i } ~ e^{-y_i} }{\Gamma(x_i + 1)},
\end{equation}
where $x_i$ and $y_i$ are the measured and expected normalized rates in the $i$-th bin, respectively, and $\Gamma(x)$ is the Euler Gamma function that generalizes the factorial for $x\in \mathbb{R}$ to a continuous variable.

Figure \ref{fig:sens} shows the median sensitivity for the expected power spectrum at one cycle/year obtained from toy Monte Carlo pseudo-experiments generated with and without the expected signal over a Borexino-like time series event rate.

Figure \ref{fig:d30} (\emph{Top}) shows the time series of the Borexino rate in the RoI in time bins of 30 days. The figure clearly shows secular trends in $R(t)$, which could bias the measured amplitude of periodic modulations \cite{Buttazzo:2020bto}. A detrending procedure is thus carried out by subtracting an empirical 
combination of exponential trends:
\begin{equation}
    R(t) = R_A e^{-t/\tau_A} + R_B e^{-t/\tau_B} \approx R_A e^{-t/\tau_A} + 
    R_B\left(1 - \frac{t}{\tau_B}\right),   
\end{equation}
where $R_A$, $R_B$, $\tau_A$, and $\tau_B$ are free parameters. The last approximation holds because $\tau_B$ is visibly much larger than the length of the data set. 
The faster decay is associated with leakage of alpha events through the MLP as well as with \pb mixing. The slower decay includes the slowly varying \bi and, possibly, $^{85}$Kr backgrounds, as discussed in Sec. \ref{sec:data}.
Figure \ref{fig:d30} (\emph{Bottom}) shows the residual rate after the detrending subtraction. The blue curve is a sinusoidal fit showing a clear annual modulation present in the time series. Details of this particular fit in relationship with the Earth's orbital parameters is described in Sec.~\ref{sec:epsilon}. 

\begin{figure}[h!]
    \centering
    \includegraphics[width=0.94\columnwidth]{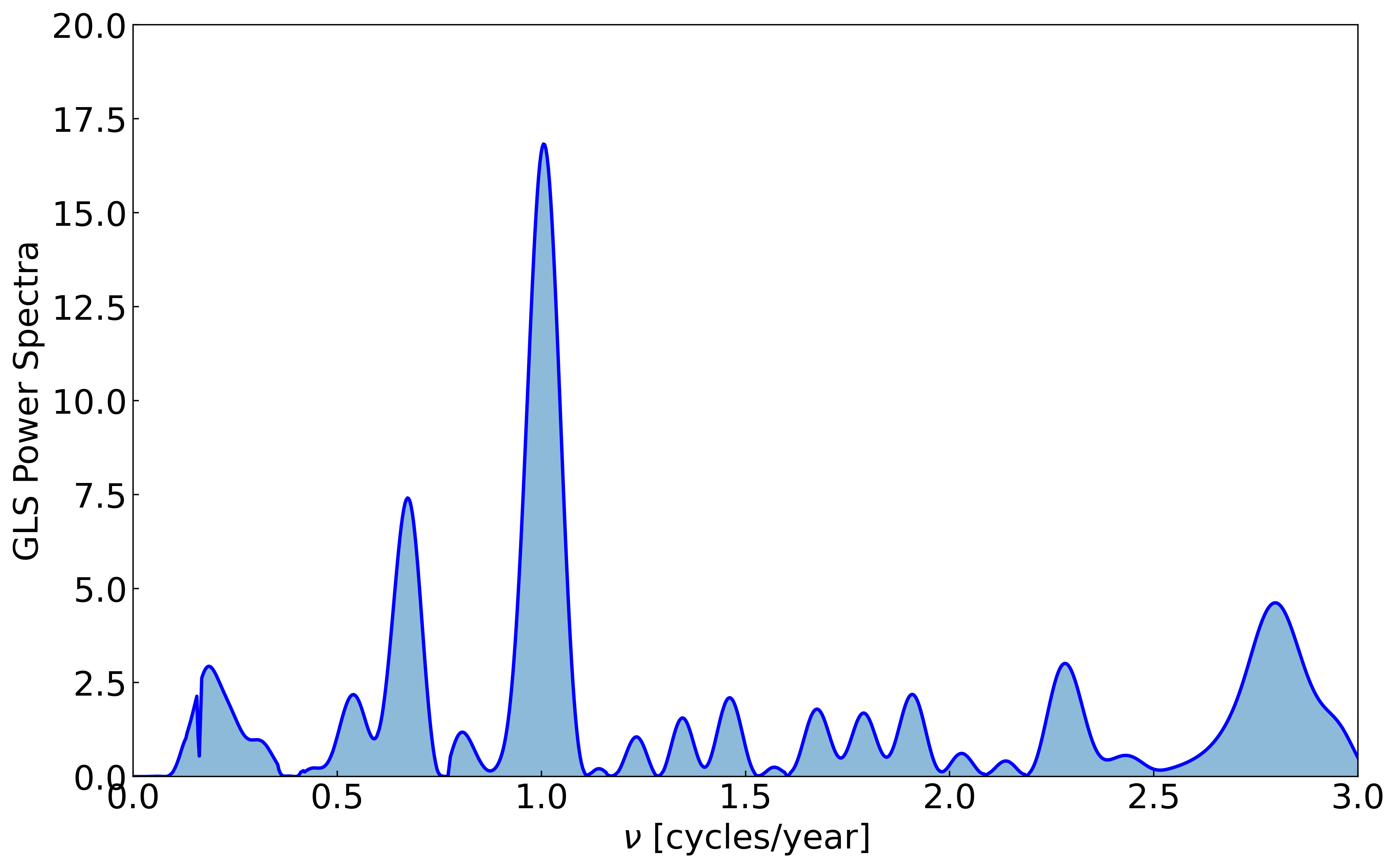}
    \caption{GLS power spectrum in $\Delta\chi^2$ units of the residual rate of Fig. \ref{fig:d30} (\emph{Bottom}). Frequencies are reported as cycles/year. A clear peak at one cycle/year frequency emerges from the full periodogram.}
    \label{fig:ps30}
\end{figure}

Finally, Fig. \ref{fig:ps30} shows the GLS periodogram  obtained from the residuals shown in Fig. \ref{fig:d30} (\emph{Bottom}). Frequencies are reported in terms of number of cycles per year (cycles/year), equal to $2.73 \times 10^{-3}$ cycles/day.
A significant peak with GLS power of 16.4 corresponding to one cycle/year frequency is clearly visible in the periodogram. 
It is noted that the frequency definition uses the \emph{anomalistic} year of 365.2596 days, defined as the time taken by the Earth to complete one revolution with respect to its perihelion. 
Considering the null hypothesis distribution $e^{-\mathcal{S}}$, the $p$-value of the peak is $5.9 \times 10^{-8}$ corresponding to $ 5.3 \sigma$ significance using the one-sided Gaussian distribution. The validity of the analytical formula for the estimation of the $p$-value was verified with a toy Monte Carlo simulation containing up to 30 million pseudo-experiments generated with pure white-noise. The absence of realizations above the measured GLS power at 1 cycle/year confirms the reported significance at more than 99\% CL.  

Figure \ref{fig:ps30} shows a second prominent peak around 0.7 cycles/year with GLS power of 7.5. At face value, the significance of this peak would be $\sim 3\sigma$ for a modulation at an expected frequency. When considering the so-called \emph{Look-elsewhere effect} (LEE), the actual significance drops to 1.8$\sigma$ (see Sec. \ref{sec:freq} for further details).

In summary, the detection of a very significant seasonally modulated component provides an independent verification of the stability of the \bi background component, confirming its large degree of uniformity already quantified in Ref.~\cite{bib:cno}.
It should be noted that the overall detector stability in terms of resolution, energy scale, and selection cuts has been amply 
corroborated over the whole Phase-II and Phase-III period.
In the following Section a detailed analysis of this periodic signal and its relationship with the Earth's orbital eccentricity is discussed, along with possible sources of systematic uncertainties.  

\section{Earth's orbit parameters}
\label{sec:epsilon}

\begin{figure}[h!]
    \centering
    \includegraphics[width=0.94\columnwidth]{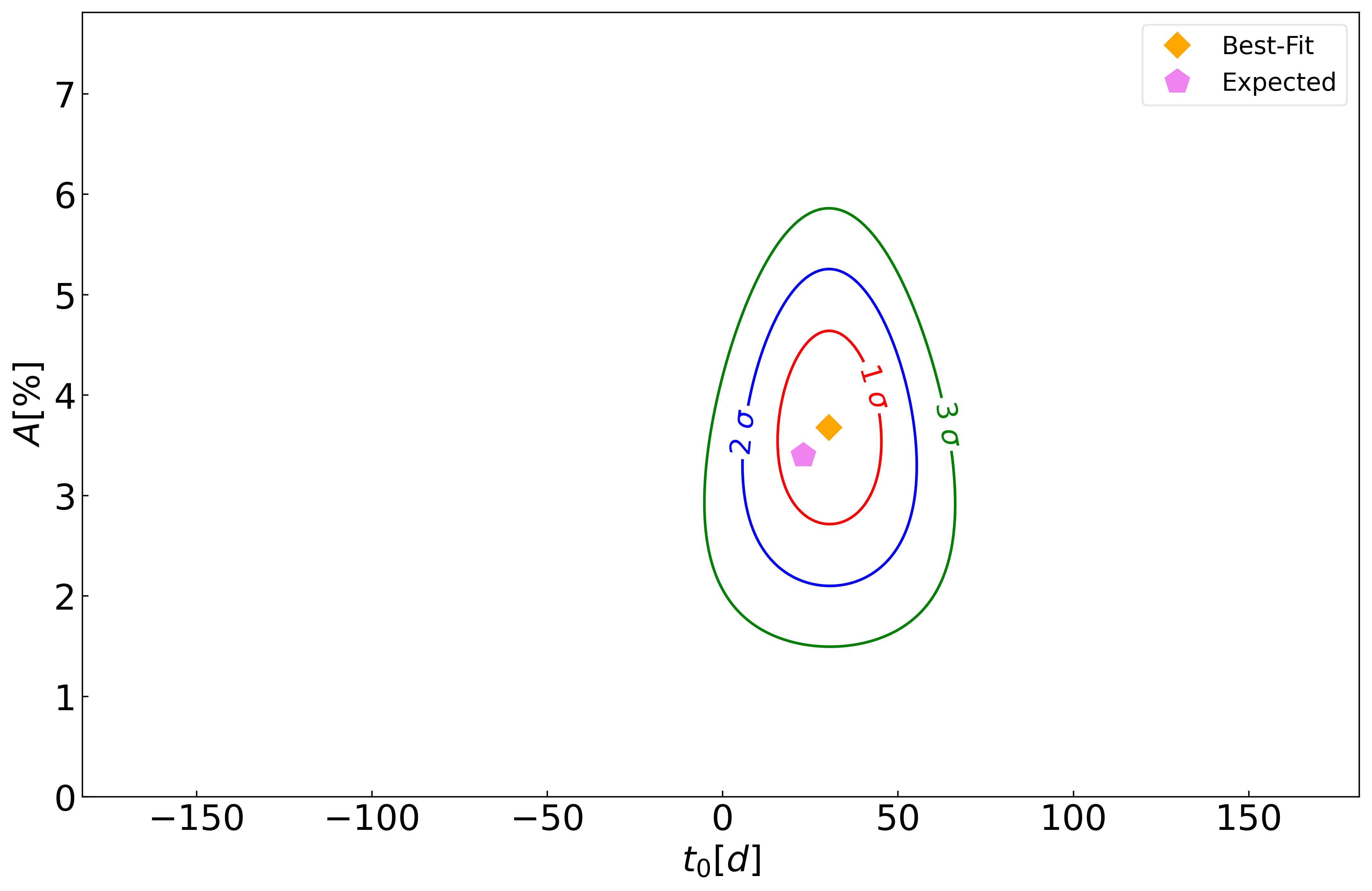}
    \includegraphics[width=0.94\columnwidth]{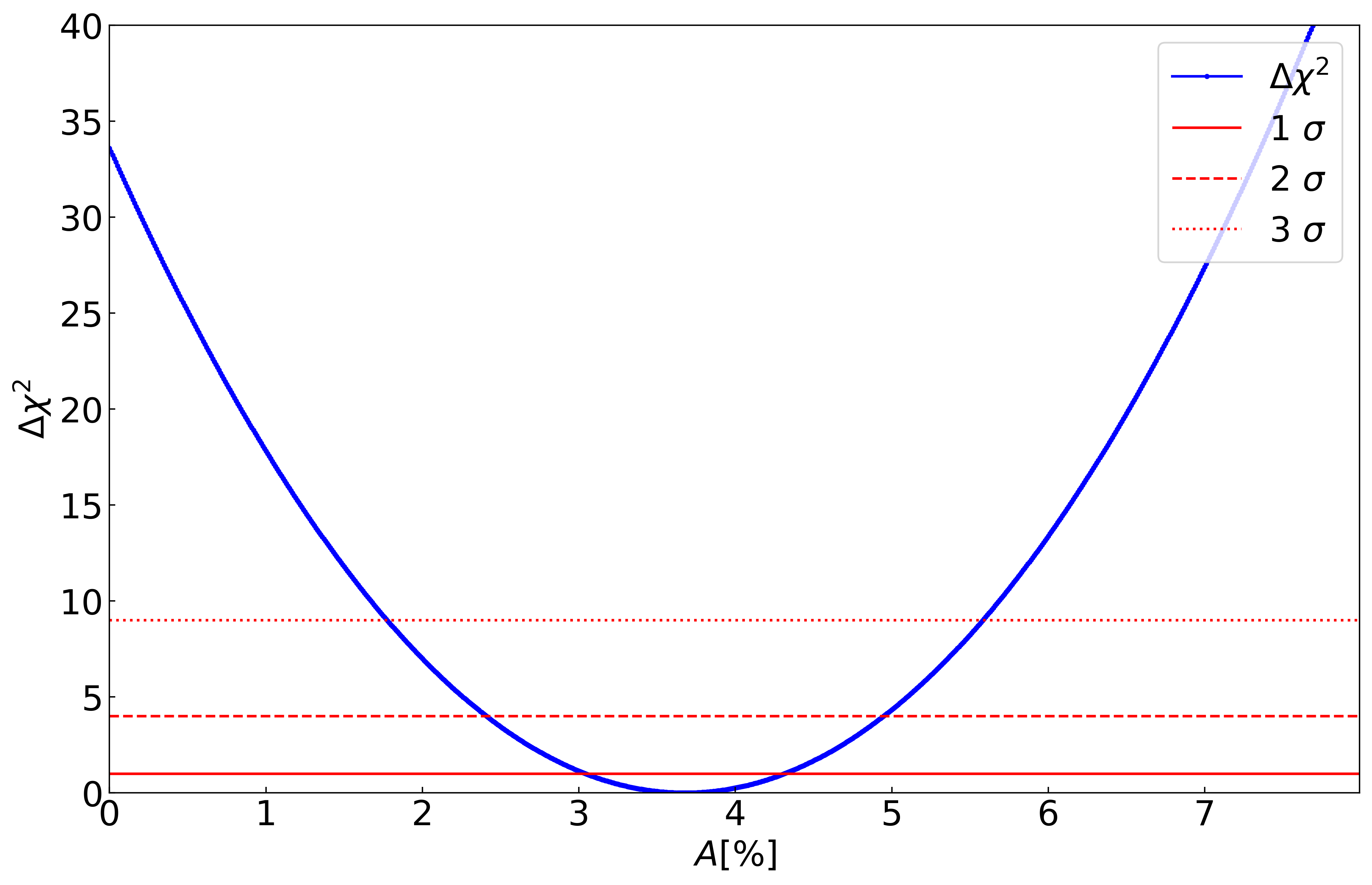}
    \caption{\emph{Top:} Best fit value (orange) and standard confidence contours for the percentage amplitude of the orbital modulation and perihelion phase from the sinusoidal fit with all free parameters. 
    The astronomical prediction is also shown (pink). \emph{Bottom:} $\Delta\chi^2$ profile over the percentage amplitude showing a significance better than $5\sigma$.}
    \label{fig:fit}
\end{figure}

To investigate the annual modulation, the time series residuals are fitted to the simple model:
\begin{equation}
    S_y(t) = A_y \cos(\omega_y(t - t_0)),
\end{equation}
where $A_y$ is proportional to twice 
the orbit's eccentricity $\epsilon$, $t_0$ is the phase shift (perihelion date) in days, and $\omega_y = 2\pi/T_y$ is the frequency, with $T_y$ nominally one year. 
$A_y, T_y$, and $t_0$ are free parameters of the fit, which  returns $A_y = (0.94 \pm 0.16) $ \cpd, $T_y = (363.1 \pm 3.6)$ days and $t_0 = (30 \pm 20)$ days, with reduced $\chi^2$ 
of $ 0.96$.
From 
Eq. \ref{eq:flux}, the flux modulation parameter is:
\begin{equation}
  A =  2\epsilon = \frac{A_y}{R_\odot} = (3.68 \pm 0.65)\%, 
\end{equation}
where $R_\odot = 25.6\pm 1.27$ \cpd is the average unmodulated solar neutrino rate, fixed to the Solar Standard model prediction, inclusive of all the model uncertainties. Notice that the uncertainty of this average rate is much larger than the precision solar neutrino flux measured by Borexino \cite{bib:global, bib:cno}. This is a conservative choice, independent of previous Borexino measurements  made on the same integrated data set. Figure \ref{fig:fit} shows the $\Delta\chi^2$ profile for the solar neutrino modulation (\textit{Bottom}) and the best fit values for the amplitude and the phase with standard confidence contours (\textit{Top}), compared with the expected values from astronomical measurement \textit{i.e.} $A = 3.37\%$ and $t_0 = 23 $ days. The latter is the perihelion date with respect to the origin of the time axis set to 12:00 AM of December 11\textsuperscript{th} 2011, in UTC time. The excellent agreement with the expected values supports the Earth's orbital origin of the annual modulation in the Borexino total rate time series. In particular, this is the first 1\%-level measurement of the annual periodicity obtained from solar neutrinos. No other significant minimum of the $\chi^2$ function is found within an annual cycle.  

The eccentricity calculated from the time fit is $\epsilon = 0.0184 \pm 0.0032$, the accuracy of the measurement is better than 20\%, and the null hypothesis (no annual modulation) is rejected at $5.9 \sigma$, as given by the intercept of the $\Delta\chi^2$ profile. 
Systematic uncertainties due to the energy scale stability, the detector resolution, the fiducial volume and data selection criteria are of the order of a few percent and, therefore, negligible. Systematic uncertainty introduced by the detrending model is also negligible. Indeed, the stability of the final results is not affected by the use of a polynomial model or a 1$^{st}$ order local regression method.

Subtler systematic uncertainties could arise from the MLP alpha removal for two main reasons. First, the MLP leakage described in Sec. \ref{sec:data} is time dependent, because of the degrading energy resolution due to PMT loss during more than 10 years of operation. To quantify the MLP alpha leakage, the full data set is split into one-year time intervals. A complementary fit of the energy spectra before and after the MLP selection in each yearly bin shows an inefficiency of $\sim$1\% at the beginning of Phase-II which grows almost linearly to 3\% by the end of data taking. Applying this inefficiency trend over the \po activity, a residual \po rate after the selection cut was studied with GLS and sinusoidal fit. A residual modulation of 0.02 $\pm$ 0.02 \cpd with a significance $<$1.1 $\sigma$ is found which, nonetheless, has no impact on the magnitude and phase of the presented results.
Second, it is known from the CNO analysis and from the thermal stabilization campaign \cite{bib:cno}, that the migration component of the \po from the Inner Vessel into the analysis fiducial volume is time dependent. Indeed, a periodic injection of \po into the detector center driven by seasonal temperature changes in the experimental hall, is observed for most of Phase-II, especially before the thermal insulation of the detector. This modulation is usually peaked around spring or early summer, thus out of phase with respect to the annual modulation expected from the Earth's orbit eccentricity. 
Finally, cosmogenic $^{11}C$ is expected to have a modest seasonal signature due to the periodic $1.4\%$ amplitude modulation of the muon flux peaked in early July \cite{bib:muons}. This effect has no measurable effect on our result since the contribution of $^{11}C$ $\beta^+$ events in the selected energy RoI is negligible.

\begin{figure}[h!]
    \centering
    \includegraphics[width=0.94\columnwidth]{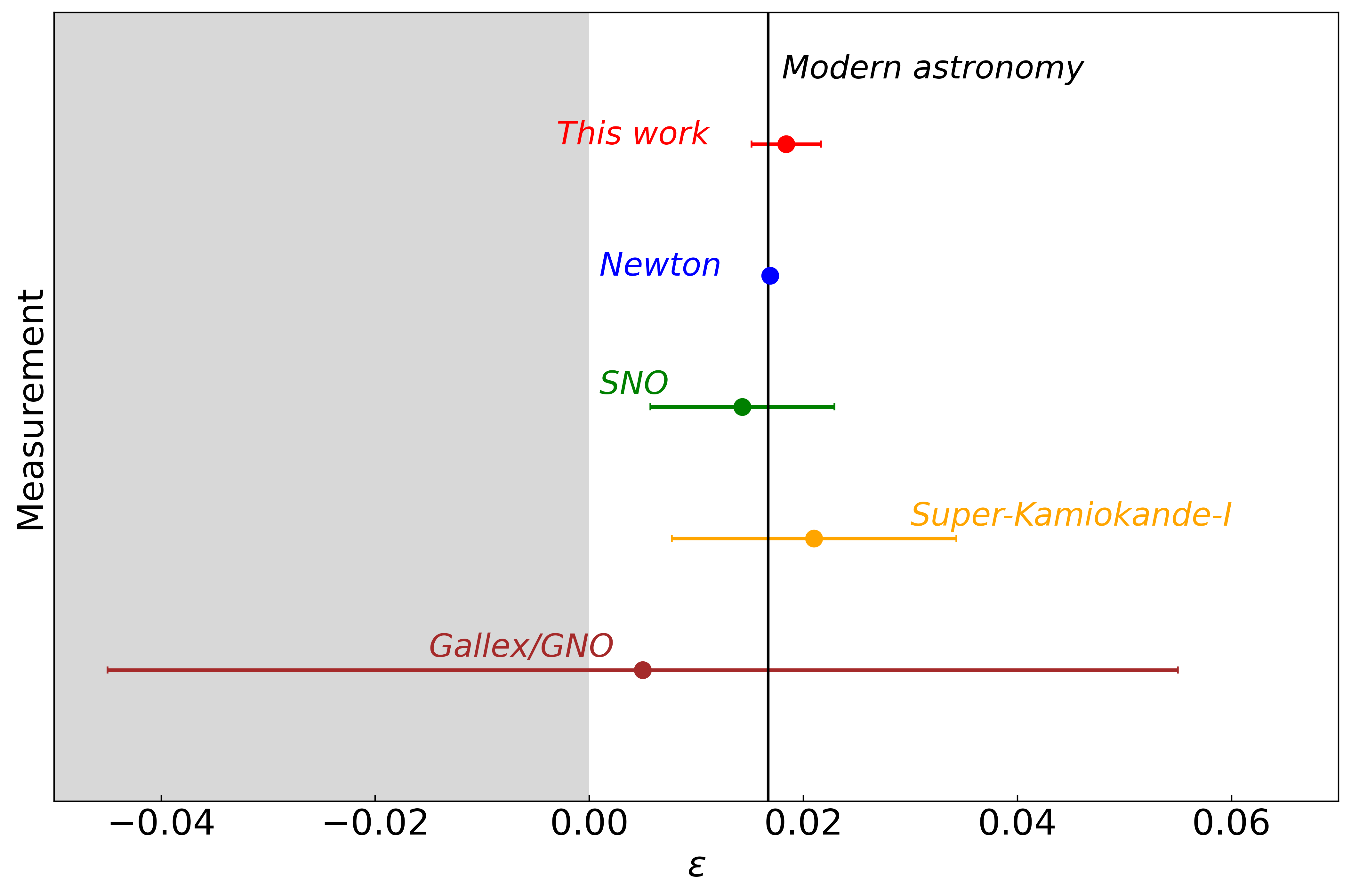}
    \caption{Comparison between the Borexino measurement of the Earth's orbital eccentricity (red) with those from previous solar neutrino experiments: SNO (green), Super-Kamiokande (yellow), and Gallex/GNO (brown)~\cite{bib:epsilon_SNO, bib:epsilon_SKI,bib:epsilon_GALLEX}. The blue point is the value reported in Newton's \emph{Principia} and the vertical black line is the current precision astronomical measurement. 
    The gray shaded region of negative values corresponds to a $\pi$ phase shift. }
    \label{fig:eccomp}
\end{figure}
Figure \ref{fig:eccomp} shows the comparison of the Borexino eccentricity measurement with those of other solar neutrino experiments. The SNO (green point) \cite{bib:epsilon_SNO} and Super-Kamiokande (yellow point) \cite{bib:epsilon_SKI} experiments searched for the annual modulation of $^8$B neutrinos selected with higher threshold (several MeV) and yielding a lower counting rate ($\sim 10$ cpd) than Borexino ($\sim 30$ cpd in the RoI, $\sim 300$ keV threshold). Evidence for annual modulation is found with 1-$2\sigma$ significance by both experiments. Gallex/GNO  (brown point)\cite{bib:epsilon_GALLEX}, set an upper limit on the modulation of the low-threshold integrated solar neutrino capture rate on gallium nuclei as it had limited sensitivity due to low event rate (order 1~\cpd).
Fig. \ref{fig:eccomp} also shows the vertical black line corresponding to the astronomical measurement with negligible uncertainty (vertical black line), the eccentricity value reported in Newton's \emph{Pincipia} (blue point), and the Borexino results (this work, red point). The gray shaded region of negative eccentricity values in the Figure corresponds to a $\pi$ phase shift. It's worth noticing that the Earth's orbit eccentricity undergoes slow secular variations classified among the so-called Milankovitch cycles. These small variations are negligible over time intervals of a few centuries and do not spoil 
as proven by 
the agreement between Newton's eccentricity value and the present astronomical measurement. See \cite{bib:nasa_e} for further details. 

Interestingly, one could derive the solar neutrino flux on Earth from the measured rate in the RoI, dominated by \be solar neutrino-electron scattering, and the eccentricity value from modern astronomy. If one neglects the contributions from $pep$ and $CNO$ solar neutrinos, assumes a relative amplitude of the modulation ($2\epsilon$) of $3.37$\%, the measured \be neutrino interaction rate (49\% of which falls within the RoI) would be $55 \pm 9$ \cpd, in good agreement with the precision value reported in Tab.~\ref{tab:solars} and with Solar Standard Model predictions. 
This result excludes the null hypothesis with $>5\sigma$ significance. In other words, Borexino could have discovered \be mono-energetic solar neutrinos via the detection of their annual modulation only, even if the characteristic \be Compton shoulder had not been visible due, {\it e.g.}, to a higher contamination of $^{238}$U and $^{232}$Th that scintillator purification could not abate.

The presence of an annual modulation in the Borexino $\beta$-like spectrum thus provides clear indication of the solar origin of a significant portion of its events.
The measurement reported here is the first precise measurement of the Earth's orbital parameters obtained solely with solar neutrinos and confirms the high stability achieved by Borexino in the last 10 year of data taking.

\section{Full periodogram analysis}
\label{sec:freq}

\begin{figure*}[h!]
    \centering
    \includegraphics[width=0.94\textwidth]{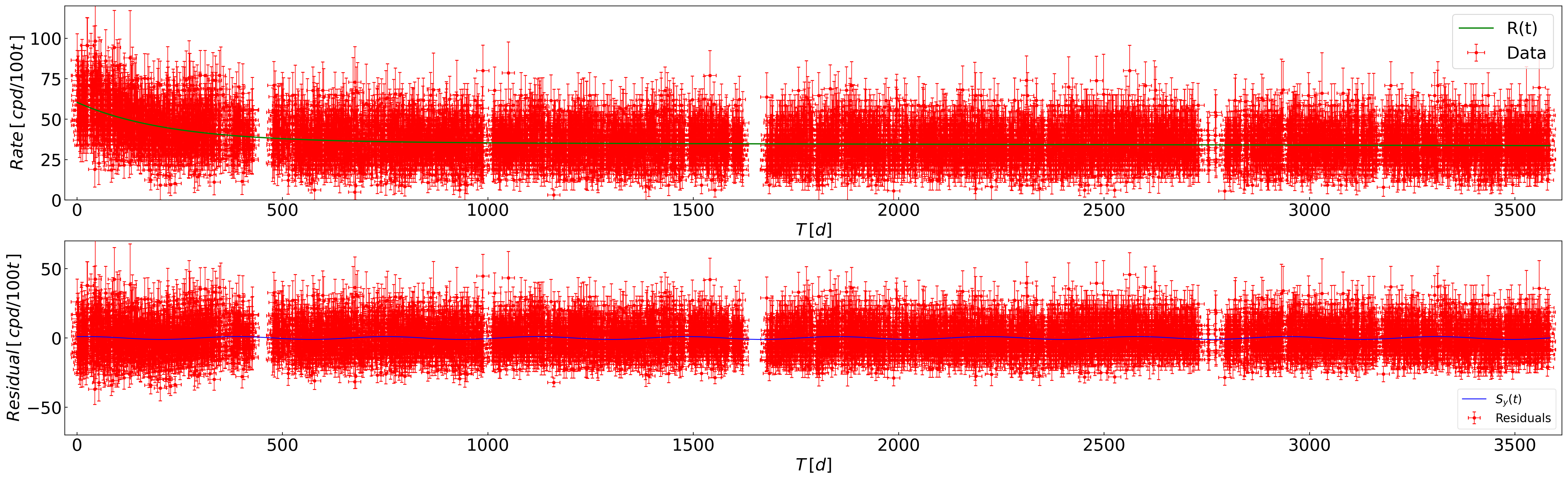}
    \caption{\emph{Top}: Full Borexino rate time series (Phase-II and Phase-III) in the RoI fitted to the trend model $R(t)$ to remove the secular components. The rate in \cpd is binned in time intervals of 8 hours. \emph{Bottom}: Residuals of the time series with respect to the trend model. The blue curve shows the sinusoidal fit of the residual rate, indicating the presence of a significant annual modulation signal. }
    \label{fig:d33}
\end{figure*}

The periodogram of all frequencies between 1 cycle/year and 547 cycles/year were studied by arranging the time series of the Borexino measured rate in time intervals of 8 hours. Figure \ref{fig:d33} shows the modified time series of the Borexino total rate in the RoI in the same analysis period.
Discontinuities in time series correspond to data acquisition breaks due to technical reasons.
This choice is a trade-off between the diurnal frequency detection capability imposed by the Nyquist theorem and the scarceness of data due to the short time binning. This optimization resulted from a toy Monte Carlo simulations study of the capability of detecting signals as a function of the selected time bin width.

\begin{figure*}[h!]
    \centering
    \includegraphics[width=\textwidth]{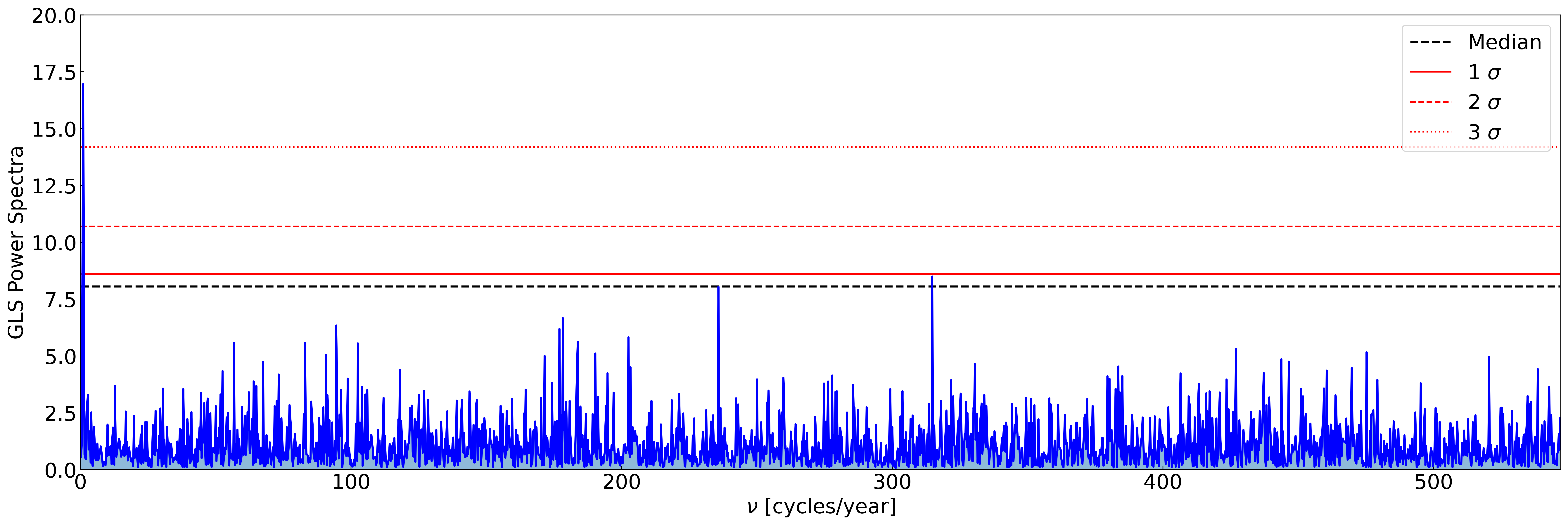}
    \caption{GLS power spectrum in $\Delta\chi^2$ units of the residual rate of Fig. \ref{fig:d33} (\emph{Bottom}). Frequencies are reported in cycles/year.  
    The median significance threshold from LEE is shown by a horizontal black dashed line, along with 1, 2 and 3 $\sigma$ significance levels (solid, dashed and dotted red lines, respectively).
    No significant peaks besides the annual modulation are present above random fluctuations of the statistical Poissonian noise.
    }
    \label{fig:ps33}
\end{figure*}

The GLS power spectrum, performed after the detrending procedure illustrated in Sec.~\ref{sec:gls}, is shown in Fig.~\ref{fig:ps33}. The annual modulation peak appears clearly on the left with significance comparable to that from a re-binning procedure. 
The significance of the other peaks was evaluated using LEE via toy Monte Carlo pseudo-experiments of white noise only whose fluctuations can randomly generate peaks in the observed frequency band. Their $p$-value distribution defines the median significance threshold shown by a horizontal black dashed line, along with the 1, 2 and 3 $\sigma$ significance levels (solid, dashed, and dotted red lines, respectively).
The LEE assigns the correct significance to random frequencies where no signal is expected. Instead, for expected frequencies, the significance of the signal is directly inferred from the normalised GLS spectrum according to the aforementioned $e^{-\mathcal{S}}$ law. 

\begin{figure}[h!]
    \centering
    \includegraphics[width=0.94\columnwidth]{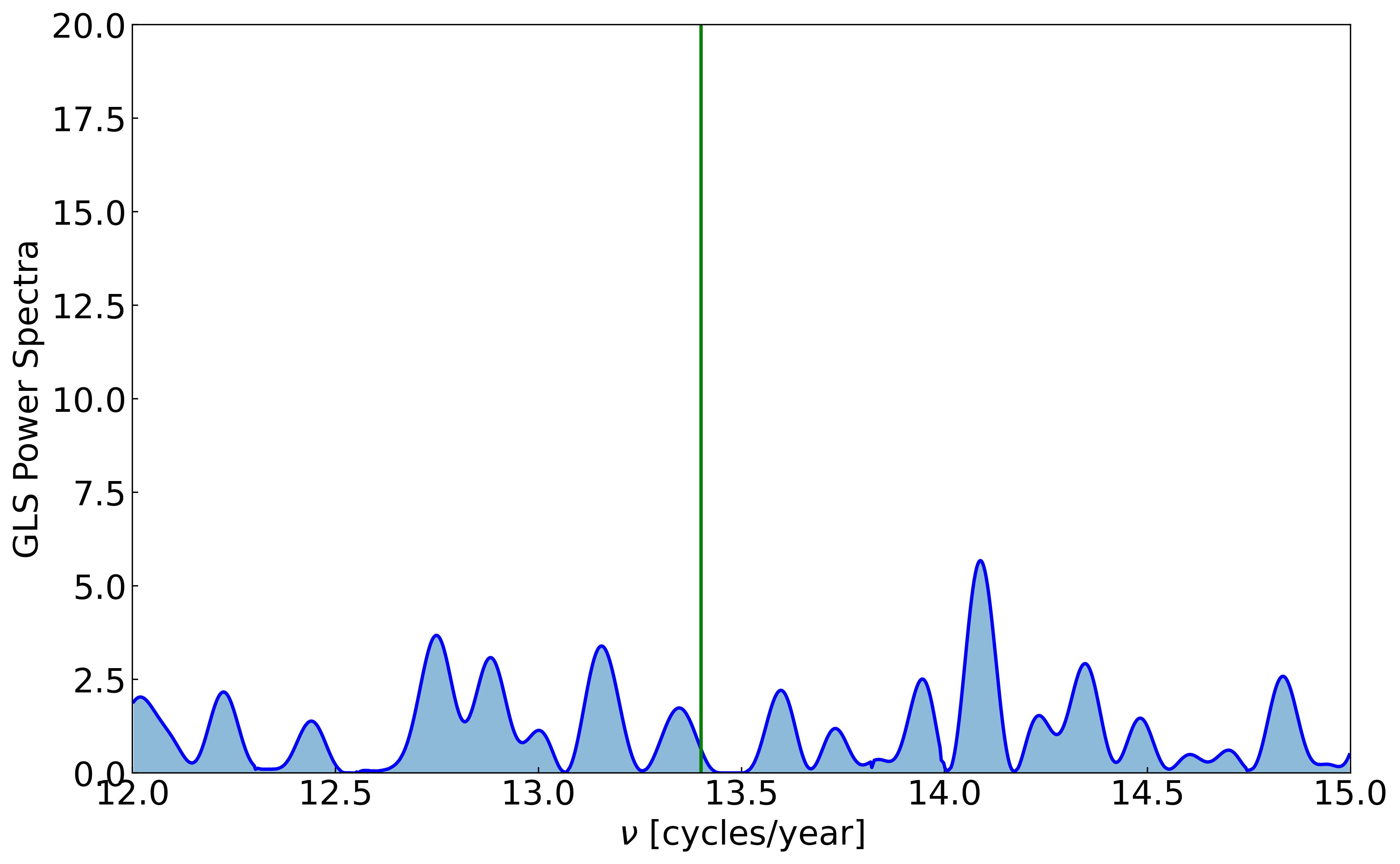}
    \includegraphics[width=0.94\columnwidth]{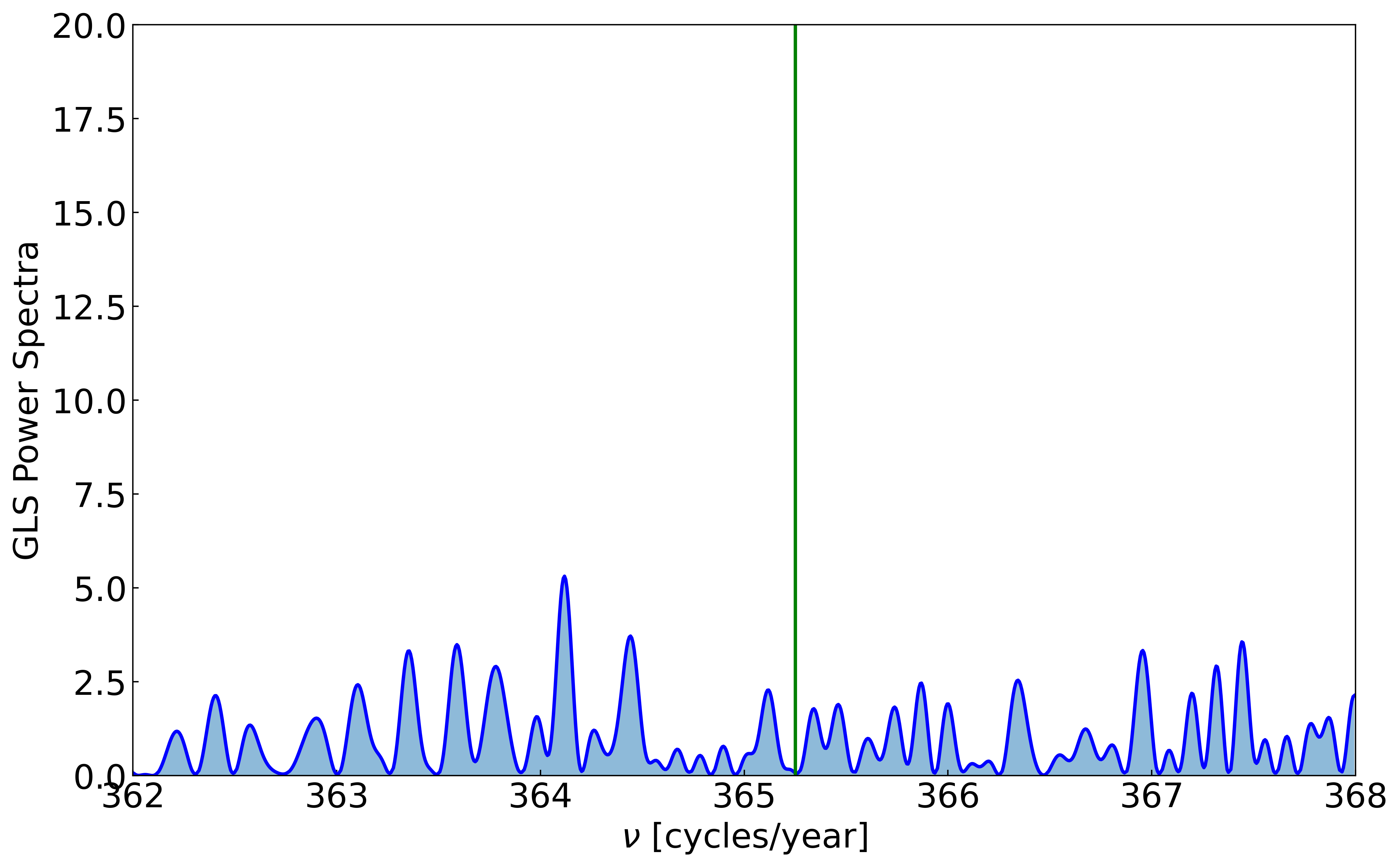}
    \caption{\emph{Top:} A zoom-in of the GLS power spectrum (in arbitrary units) of the residual rate of Fig. \ref{fig:d33} around the one cycle/month point. The Sun's synodic rotation line of $13.4$ cycles/year is shown (green vertical line). \emph{Bottom:} A zoom-in of GLS power spectrum (in arbitrary units) of the residual rate of Fig. \ref{fig:d33} around the one cycle/day point. The Earth's rotation line of 365.2596 cycles/year is shown (green vertical line).}
    \label{fig:ps33zoom}
\end{figure}

Figure \ref{fig:ps33zoom} shows zoom-ins of the GLS spectrum in the one cycle/month range, {\it i.e.}, around the Sun's Synodic Carrington rotation frequency of 13.4 cycle/year (\emph{Top}), and around the diurnal modulation frequency of 1 cycles/day (365.2596 cycles/year, \emph{Bottom}). The $\sim$monthly frequency could reveal some anisotropy of the Sun, somehow affecting neutrino production during its axial rotation. The $\sim$daily frequency is coupled to electron neutrino regeneration in the Earth, of interest for sterile neutrino theories~\cite{bib:cirelli, ryan1, ryan2, ryan3}. Other theoretical scenarios investigated via time modulations of solar neutrinos include the search for new interactions beyond the Standard Model~\cite{bib:roul, bib:guzzo}, such as non-standard interactions (NSI) and alike~\cite{bib:gago, bib:palazzo, bib:nsi01, bib:nsi02} for recent investigations.

Borexino Phase-I tightly constrained the day-night asymmetry of the solar neutrino interaction rate, $A_{dn} = 2(D-N)/(D+N) = 0.001\pm 0.012~(stat) \pm 0.007~(syst)$,
with $D$ and $N$ the integrated day and the night rates, respectively~\cite{bib:day-night}.
For reference, the day-night variation for the monochromatic \be neutrinos (866 keV) in the standard three-flavour neutrino scenario is $A_{dn} \approx 6 \times 10^{-4}$ \cite{bib:vissani}.
To compare the Phase-I with the present analysis performed over the Phase-II+Phase-III complementary data set,
the time series residual of Fig.~\ref{fig:d33} (\emph{Bottom}) is fitted to a sinusoidal function as in Sec.~\ref{sec:gls}, $f(t)=A_d \cos(\omega_d t + \phi_d)$, 
where $A_d$ is the amplitude of the daily modulation and
both the frequency $\omega_d$ and the $\phi_d$ are fixed to one cycle/day and the local midnight, respectively
$A_{dn}$ is approximately related to $A_d$ as:
\begin{equation}
    A_{dn} = \frac{ 2A_d}{\sqrt{2} R_\odot}\,
\end{equation}
where, $R_\odot = 25.6 \pm 1.27$ \cpd (see Sec.~\ref{sec:epsilon}).
The extra $\sqrt 2$ factor comes from the integration of the sinusoidal day-night modulation over a 24 h period.
We obtain $A_{dn} = 0.0030 \pm 0.0094 (stat)  \pm 0.0002 (sys)$, compatible with 0 at $1\sigma$ level. The systematic uncertainty is dominated by the solar modeling as for the determination of the eccentricity in Sec.~\ref{sec:epsilon}. This number should not be directly compared with the Borexino Phase-I result, for which the day-night spectra were defined taking into account the seasonal variation of the duration of a day and its effect on the actual neutrino trajectory through the Earth. Specifically, the day-night effect is a superposition of a diurnal modulation with a sub-dominant annual carrier correlated with the day-night amplitude. A more accurate analysis is, however, not expected to yield drastically different results allowing us to conclude that the expected annual modulation is the only statistically significant frequency in the Borexino time series.

\section*{Conclusions}

The Borexino experiment concluded its data taking in October 2021 after more than 14 years of activity. 
The time series of the the total solar neutrino rate over the last 10 years was analyzed using events selected in a fixed energy window chosen to maximize the signal-to-background ratio. We have searched for solar neutrino signal modulations in the frequency range between one cycle/year and one cycle/day using the generalized Lomb-Scargle method.

We identified no significant periodic signal other than the annual modulation due to the Earth's orbit eccentricity. The latter is measured with amplitude (related to the orbit eccentricity), phase (perihelion position), and frequency (Earth revolution) parameters compatible within one sigma with astronomical predictions. In particular, the best-fit eccentricity is $\epsilon = 0.0184 \pm 0.0032$ (stat+sys), with the null hypothesis excluded with a significance greater than $5\sigma$. 
This results is the most precise measurement of the Earth's orbit eccentricity obtained using solar neutrinos alone.

No other significant modulation of the backgrounds is expected in the selected energy range. In particular, the critical \po leakage events, due to the inefficiency of the pulse shape discriminator, were carefully quantified as negligible. Residual, well-characterized time variations of the background are limited to slow, monotonic trends, which are easily removed by a detrending procedure.

Strong constraints are placed on the amplitudes of other frequencies of interest, {\it i.e.}, day-night effects and correlations with the Sun's rotation around its axis.
Both frequencies are not significant using the LEE approach. In particular, the limits for the percent diurnal modulation and the percent solar rotation day are $< 1.3\%$ (90\% CL) and $1.8\%$ (90\% CL), respectively.
These improved bounds are relevant in solar modelling and in constraining a wide variety of non-standard neutrino interactions beyond the Standard Model of particle physics and the present three-flavour neutrino oscillation paradigm.  

The ability of Borexino to measure the expected annual modulation of its neutrino signal further confirms its solar origin and adds to the experiment's success in measuring, with high precision, all solar neutrino fluxes emitted in the hydrogen burning processes (pp-chain and CNO cycle) in the Sun. 
This measurement was enabled by the   
stability of the detector response and energy resolution, as well as by the exquisite understanding of the radioactive background contamination of the detector.

\section*{Acknowledgments}

We thank Francesco Vissani for useful discussions about the implications of the present results for neutrino physics. We also thank former Borexino collaborator Francesco Lombardi for useful information about the annual modulation analysis.

The Borexino program is
made possible by funding from Istituto Nazionale
di Fisica Nucleare (INFN) (Italy), National Science
Foundation (NSF) (USA), Deutsche Forschungs 
gemeinschaft (DFG) and Helmholtz-Gemeinschaft
(HGF) (Germany), Russian Foundation for
Basic Research RFBR (Grant 19-02-00097 A),
RSF (Grant 21-12-00063) (Russia), and Narodowe Centrum Nauki (NCN) (Grant No. UMO
2017/26/M/ST2/00915) (Poland). This research
was supported in part by PLGrid Infrastructure.
We acknowledge the generous hospitality and
support of the Laboratory Nazionali del Gran
Sasso (Italy).

\end{document}